\documentclass[letter]{aa}  
\usepackage{graphicx}
\usepackage{hyperref}
\hypersetup{colorlinks=true,allcolors=[rgb]{0,0,1}}
\usepackage{xcolor}
\usepackage{amsmath,bm}
\usepackage{comment}
\usepackage{pdflscape}

\usepackage{txfonts}
\newcommand{\Teff}{$T_{\text{eff}}$}
\newcommand{\Lspalone} {log\,$\mathcal{L}$}

\newcommand{\He}{Y$_{\rm He}$}
\newcommand{\vsini}{$v$\,sin\,$i$}
\newcommand{\Nab}{$\epsilon_{\rm N}$}
\newcommand{\kms}{km\,s$^{-1}$}

\newcommand{\SSEMoutliers}{{SSEM-outliers}}
\newcommand{\SSEM}{{SSEM}}
\newcommand{\inprep}{\textcolor{blue}{Martínez-Sebastián et al. (in prep.)}}

\begin{document}

   \title{The IACOB project}

   \subtitle{XIII. Helium enrichment in O-type stars as a tracer of past binary interaction}

   \author{C. Martínez-Sebastián
          \inst{1,2}
          \and
          S. Simón-Díaz
          \inst{1,2}
          \and
          H. Jin
          \inst{3}
          \and
          Z. Keszthelyi
          \inst{4}
          \and
          G. Holgado
          \inst{1,2}
          \and
          N. Langer
          \inst{3,5}
          \and
          J. Puls
          \inst{6}
          }

   \institute{Instituto de Astrofísica de Canarias, c/Vía Láctea, S/N, E-38205 La Laguna, Tenerife, Spain
   \and
   Departamento de Astrofísica, Universidad de La Laguna, E-38206 La Laguna, Tenerife, Spain
   \and
   Argelander Institut für Astronomie, Auf dem Hügel 71, DE-53121 Bonn, Germany
   \and
   Center for Computational Astrophysics, Division of Science, National Astronomical Observatory of Japan, 2-21-1, Osawa, Mitaka, Tokyo 181-8588, Japan
   \and
   Max-Planck-Institut für Radioastronomie, Auf dem Hügel 69, DE-53121 Bonn, Germany
   \and
   LMU Munich, Universitätssternwarte, Scheinerstrasse 1, 81679 München, Germany}
   
   \date{Received xx,xxxx; accepted xx,xxxx}
 
  \abstract
   {There is increasing evidence that single-star evolutionary models are inadequate to reproduce all observational properties of massive stars.
   Binary interaction has emerged as a key factor in the evolution of a significant fraction of massive stars.
   In this study, we investigate the helium (\He{}) and nitrogen (\Nab{}) surface abundances in a comprehensive sample of 180 Galactic O-type stars with projected rotational velocities $\leq$150\,\kms. 
   We found a subsample ($\sim$20\,\% of the total, and $\sim$80\,\% of the stars with \He{}$\geq$\,0.12) with a \He{} and \Nab{} combined pattern unexplainable by  single-star evolution.
   We argue that the stars with anomalous surface abundance patterns are binary interaction products.
}


   \maketitle


\section{Introduction}\label{Intro}

Since the early 1970s, the reliable interpretation of surface abundance patterns of CNO-cycle products in main sequence O- and B-type stars has remained challenging. 
According to the standard theory of stellar structure and evolution, non-rotating massive stars do not bring nuclear-processed matter to their outermost radiative layers during the main sequence. Thus the composition of the stellar surface should resemble the initial abundances.
However, for more than five decades, there has been clear and continuously increasing observational evidence of enhanced helium (He) and nitrogen (N) abundances in the photospheres of a significant number of these stars \citep[see, e.g.,][]{Lester+73, Schonberner+88, Herrero+92, Lyubimkov+96, Howarth2001, Morel+06, Hunter+08, Hunter+09, Rivero-Gonzalez+12, Bouret+13, Bouret+21, Martins+15, Martins+17, Martins+24, Markova+18, Grin+17, Carneiro+19}. 

Internal mixing processes induced by stellar rotation were initially proposed as a robust theoretical explanation for the observed abundances \citep[see review by][]{Maeder&Meynet00}. However, new observations from spectroscopic surveys soon began to reveal certain limitations of this scenario \citep[e.g.,][]{Hunter+08, Hunter+09, Brott+11}, highlighting the need for additional physical mechanisms to explain the occurrence of contaminated stellar surfaces. 
At this juncture, mass transfer and merging events in binary systems were suggested to play an important role \citep[see, e.g.,][and references therein]{Langer12}, with complementary studies arguing for the potential contribution of internal gravity waves \citep{Aerts+14}.
Also, the properties of a distinct sub-population seem to be best explained via magnetic stellar evolution models \citep[e.g.,][]{Potter+12, Keszthelyi+19, Keszthelyi+22, Takahashi&Langer21}.

Since then, many objects previously thought to be single stars have been discovered to be part of binary or even higher-order multiple systems \citep[e.g.][]{Sana+14, Aldoreta+15, Barba+17, MaizApellaniz+19}. Furthermore, the majority of massive stars have been claimed to undergo interaction processes throughout their lifetime \citep{Sana+12}. Consequently, a rich set of post-interaction products are expected to be formed \citep[such as stripped stars, rapidly rotating accretors, or
merger stars; see review by][]{Marchant&Bodensteiner23}, broadening the spectrum of observational cases to be interpreted.

The increasing availability of high-quality spectroscopic observations of massive stars \citep[e.g.][]{Evans+05, Evans+11, Simon-Diaz+11b, IACOB2020, Maiz-Apellaniz+20, Villasenor+21, Shenar+24} allows for a more comprehensive study, improving parameter coverage and enhancing statistical significance. This advancement enables more efficient identification of specific subsamples, which can be used to understand better and constrain the various proposed physical mechanisms leading to contaminated surfaces in massive main sequence stars.
This is the case of the study presented in this Letter, where we provide first strong empirical evidence of the identification of a specific subset of 36 Galactic O-type stars (in a sample of 180 targets) whose surface abundance pattern of He and N cannot be produced by any of the state-of-the-art single-star evolutionary models and hence are candidates to trace a past binary interaction event.
\section{Sample and methods}\label{Methods} 
 
At present, the IACOB database\footnote{\href{http://research.iac.es/proyecto/iacob/iacobcat/}{http://research.iac.es/proyecto/iacob/iacobcat/}} \citep[last described in][]{IACOB2020} comprises 373 Galactic O-type stars not identified as a double line spectroscopic binary (SB2) or a peculiar star (e.g. Oe and Of?p). 
To allow for a more reliable N abundance analysis (see below), we concentrate on those (273) stars with a projected rotational velocity (\vsini{}) $\leq$\,150\,\kms{}.

Details about the analysis strategy followed for the determination of the line-broadening and spectroscopic parameters (including the He abundance by number, \He{}$\,=\!{\rm He}/{\rm H}$) can be found in \citet{Simon-Diaz&Herrero14} and \citet{Holgado+18}, respectively. Line-broadening parameters (\vsini\ and $v_{\rm mac}$) were directly adopted from \citet[][]{Holgado+22}, following a similar strategy for new observations. However, aiming to improve the accuracy of our estimates of the He abundances, and to minimize as much as possible the associated uncertainties, we decided to perform a new spectroscopic analysis with the IACOB-GBAT tool \citep{Simon-Diaz+11b} benefiting from an extended version of the grid of FASTWIND \citep{Santolaya-Rey+97,Puls+05} models used in \cite{Holgado+20}.
In particular, the new grid of FASTWIND (v10.6.5) models considers a reduced grid step size in \He{} from 0.05 to 0.02. It also includes a better sampling of microturbulence below $\xi_{\rm t}$\,=15~\kms, and two additional grid points at 25 and 30~\kms.

We then followed the strategy proposed in \cite{Carneiro+19}, based on a $\chi^2$ fitting of the equivalent width (EW) of the available N\,{\sc ii-v} lines, to obtain estimates for the N abundance (\Nab{}\,=\,log(${\rm N}/{\rm H}$)+12). This imposed the above-mentioned limitation in \vsini{}, as the lines in spectra with larger \vsini{} display much broader and shallower profiles, hence preventing obtaining reliable EW measurements. 
This method was preferred both for being faster and not critically sensitive to the line-broadening parameters compared to the alternative profile fitting. We used the N model atom developed and described in \cite{Rivero-Gonzalez+11} and a similar set of diagnostic lines as quoted there. We refer the reader to \inprep{} for a thorough description of the analysis strategy, the quality of the obtained results per individual stars, and some associated caveats and limitations found during the process.

From the initial sample of 273 stars with \vsini\,$\leq$\,150\,\kms, we could obtain reliable N and He abundances for 180 of them. Among the remaining targets, about half of them were discarded due to either (1) the low quality of the available spectra, in terms of signal-to-noise ratio, which prevented accurate measurement of the EW of the main N diagnostic lines, or (2) the absence of sufficiently strong N lines in the spectrum. The other half were eliminated from our study because the quality of the {\sc IACOB-GBAT} outcomes was insufficient to provide reliable estimations of stellar parameters.

To facilitate the interpretation of the results, we also incorporate information about the spectroscopic binarity and runaway status into our study. For the former, we benefited from the multi-epoch character of the IACOB database, which includes a minimum of 3 spectra for more than 75\% of the whole sample of O-type stars. This helped us to separate the sample into single-line spectroscopic binaries (SB1) and likely single (LS) stars. To this end, we followed the guidelines presented in \citet{Holgado+18} and \citet{Simon-Diaz+24}. The runaway status was extracted from \cite{Maiz-Apellaniz+18} and \cite{Carretero-Castrillo+23}, both based on information about proper motions as delivered by the {\em Gaia} mission.

\section{Results}\label{Results} 

\begin{figure*}
\begin{center}
\resizebox{0.99\hsize}{!}{\includegraphics{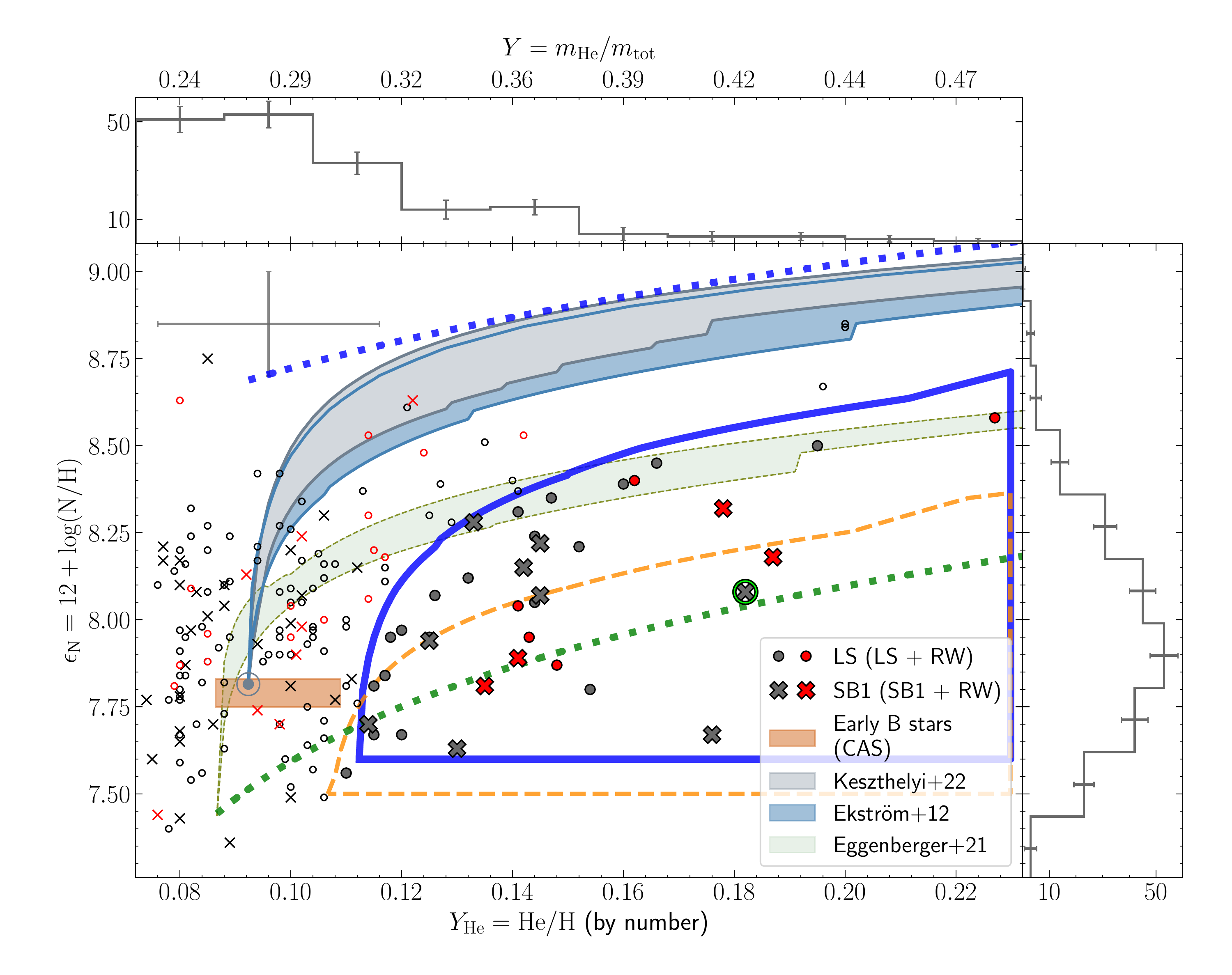}}
  \caption{ 
  He abundance (in number fraction in the lower x-axis, and in mass fraction in the upper one) against N abundances, for a sample of 180 Galactic O-type stars with \vsini\,$\leq$150~\kms. Circles and crosses indicate LS and SB1 stars, respectively, while identified runaway stars are highlighted by red symbols. Typical uncertainties in He and N abundances are marked with a gray errorbar in the upper left corner. HD\,226868 (optical counterpart of Cyg X-1) is surrounded by a green circle. Gray and blue shadowed strips embrace the location of Solar-abundances single star evolution tracks associated with GENEC $v_{\rm ini}/v_{\rm crit}$\,=\,0.2 and 0.4 model computations \cite[][and priv. comm.]{Ekstrom+12}, and MESA models following \cite{Keszthelyi+22}. Baseline abundances for these models \citep[from][]{Ekstrom+12} are marked with a bluish-gray dot. The green shadowed strip embraces the location of single star evolution tracks associated with GENEC $v_{\rm ini}/v_{\rm crit}$\,=\,0.4 model computations for subsolar initial chemical composition \citep[][see further discussion in Appendix~\ref{subsolar}]{Eggenberger+21}.
  Dotted lines correspond to two limit cases of matter mixing at the stellar surface (see Appendix~\ref{app_acc}).
  The blue-dotted line corresponds to mixing in which CNO-equilibrium is reached for any \He{} value \citep[considered initial values from][]{Ekstrom+12}. The green-dotted line corresponds to the mixing of pure He with material in CNO-equilibrium \citep[considered initial values from][see the justification for this choice in the text]{Eggenberger+21}.
  The shaded brown rectangle indicates the Cosmic Abundance Standard obtained for early-B stars in the Solar vicinity by \cite{Nieva&Pryzbilla12}. The blue open contour surrounds the star sample with N and He abundance patterns not covered by the Solar-abundances evolutionary models; the same for the orange open contour (with dashed line) but for subsolar-abundances evolutionary models. Further details on the rationale for including the subsolar-abundances tracks can be found in Appendix~\ref{appSingle}. 
  {\it Top and right:} Histograms of He and N abundances, respectively.
  \label{sum}
}
\end{center}
\end{figure*}

Figure~\ref{sum} summarizes the key findings of our study concerning the He and N abundance analysis, as well as the identification of spectroscopic binaries and runaway stars within the targeted sample.
Approximately 25\% of the total sample falls within the high He regime (\He{} $\geq 0.12$). Notably, $\sim$80\% of them (36 stars, highlighted with larger filled symbols and surrounded by a blue contour) exhibit unexpectedly low N abundances that are inconsistent with the He/N ratio predicted by the CNO cycle (see Sect.~\ref{evolpred}).
This group is particularly remarkable when compared to previous observational studies that employed similar methods and the same analysis tools (though concentrating on smaller stellar samples), but missed the identification of a significant subpopulation with the abovementioned characteristics (see Fig.~8 in \citeauthor{Rivero-Gonzalez+12} \citeyear{Rivero-Gonzalez+12} and Fig.~13 in \citeauthor{Grin+17} \citeyear{Grin+17}), presumably due to the smaller sample size.
This letter focuses on this subsample (detailed in Table~\ref{BinList_table}, along with relevant information), and the origin of the detected "anomalies". Hereafter, we refer to these stars as \SSEMoutliers{}, where \SSEM{} stands for single star evolution models.
The predictions from such \SSEM{}, together with other important information of interest, have been added to Fig.~\ref{sum}:
The paths followed by the three sets of \SSEM{} described in Appendix~\ref{appMESA} (two for solar initial abundances, and a third one assuming subsolar initial values) are shown with the shaded gray, blue, and green areas, respectively.
Only the portions of the tracks corresponding to core hydrogen burning (i.e., main sequence evolution) are considered. 
Interestingly, and regardless of different physical assumptions used in the various computations -- particularly concerning mixing mechanisms, angular momentum transport, and initial rotational velocities (see Appendix~\ref{appMESA}) -- all models follow relatively tight tracks on the diagram, similar for similar initial abundances. 
This is a fundamental prediction in the single stellar evolution of massive stars, independently of the model considered, as it is the result of the CNO cycle. 
Remarkably, the \SSEMoutliers{} do not fit into this fundamental prediction.

The shaded brown rectangle at the bottom left of the figure indicates the present-day cosmic (He and N) abundance standard suggested by \citet{Nieva&Pryzbilla12}, resulting from the analysis of a sample of several tens of main sequence early B-type stars in the Solar neighborhood (covering distances from the Sun up to $\sim$500\,pc). This area of the diagram can be considered representative of the baseline abundances of our investigated sample of O-type stars, and it roughly coincides with the bluish-gray circle, which indicates the initial abundances assumed in the evolutionary models for solar metallicity.

Histograms of He and N abundances are shown in the top and rightmost panels. Error bars in the histograms were computed following a Markov Chains Monte Carlo (MCMC) approach, considering the individual uncertainties associated with the He and N abundance estimates. These uncertainties range between the 20\,--\,30\% in the case of He, and are typically on the order of $\sim$\,0.15\,dex for the N abundances (see gray cross in the upper left of the main panel, and further notes in Sect.~\ref{obs_errors}).

Despite the significant fraction of \SSEMoutliers{}, the majority of the analyzed stars ($\sim$75\% of the sample) have normal-to-low He abundances (\He{}\,$<$\,0.12), where those objects exhibiting low values (20\% of the sample with \He{}\,$<$\,0.09, still consistent with normal values considering typical uncertainties) will be further discussed in a forthcoming study. Nitrogen, on the other hand, appears to be approximately normally distributed, around 
\Nab{}\,$\sim$7.9~dex, with only few extremely N-enriched objects.

\section{Discussion}\label{Disc}
Single-star evolution models fail to explain the outliers found on the N-He plane (He enriched but N unenriched stars). Observational biases, lower initial abundances due to the present-day Galactic abundance gradient, or other unconsidered evolutionary pathways are also unlikely explanations (see Appendix~\ref{appSingle} for a detailed discussion). Consequently, post-binary interaction rises as the most compelling origin for these objects.

\subsection{Evolutionary model predictions}
\label{evolpred}

In single stars, internal mixing and wind stripping can make nuclear-processed material appear at the stellar surface. CNO equilibrium is achieved in the convective core near the beginning of the main sequence (MS) evolution, before a significant mean molecular weight gradient develops at the core-envelope interface and before significant He-enrichment in the core. Helium, on the other hand, becomes synthesized in the core only on the nuclear timescale, and mixing it into the envelope is hindered by the corresponding mean molecular weight gradient. 
Also, mass loss is uncovering the layers with the least He-enrichment -- but still very N-enriched -- first. 
This is why N-enhancement approximately precedes He-enhancement in single stars (self-enrichment).

For mass gainers in massive binary systems, the situation is different. During the mass transfer, at first, matter that is not much enriched or even pristine is moved to the mass gainer. Towards the end, almost pure He is transferred, including N at its CNO-equilibrium value.
The mixing of this matter with the envelope of a sufficiently massive gainer with an unmixed atmosphere -- mainly composed by pristine gas -- naturally leads to significant surface He enrichments with only modest N enhancements. 
This is shown by the comprehensive, detailed binary evolution models of \citet[][see also Appendix~\ref{app_acc}]{Jin+24,Jinprep}.

As further described in Appendix~\ref{app_acc}, binary interaction products can cover a much broader range in the \He{}-\Nab{} plane compared to \SSEM{} with the same initial composition. This is illustrated in Fig.~\ref{sum} by the two dotted lines representing two extreme cases in which mixing of processed material reaching the stellar surface can occur. The upper limit (in blue) corresponds to the case in which CNO-equilibrium is reached for any \He{} value, with a CNO baseline consistent with solar values (second column of Table~\ref{in_ab}). The lower boundary (in green) considers the mixing of gas with the initial chemical composition with pure He matter in CNO equilibrium, where a subsolar CNO baseline is assumed (third column of Table~\ref{in_ab}). The latter was chosen to be consistent with stars with the lowest N abundance in our sample and to account for the possibility that some of the stars in our sample were born in regions with a subsolar composition (see Appendix~\ref{subsolar}).
Remarkably, most of the \SSEMoutliers{} lie, within uncertainties, within the region limited by the two dotted lines. The only two targets that clearly remain below the green dotted line have been identified as SB1, where the dilution by contamination from the unseen companion could have affected the nitrogen determination.

All considered, stars identified as \SSEMoutliers{} have likely not evolved in isolation, but would have been gainers in a binary system following a Roche Lobe Overflow (RLOF) mass transfer event.

\subsection{Spectroscopic binaries and runaways}

The binary origin of these systems is expected to result in two distinct dynamical outcomes following the mass transfer event \citep[see Fig.~4 and associated text in][]{Marchant&Bodensteiner23}. First, when the primary star completes its nuclear burning, it may undergo a supernova explosion, potentially disrupting the system and creating a runaway star if the kick's direction and momentum are favorable \citep[][and reference therein]{Blaauw61,van_den_Heuvel81}. Given the short post-mass transfer lifetime of the primary (as it has already left the main sequence) and a comparable rate of dynamical ejections associated with the star formation process for any type of single or multiple systems \citep{Poveda+67}, this suggests a higher fraction of runaways among the \SSEMoutliers{}.

Alternatively, if the system remains bound, the primary star is expected to lose its hydrogen envelope before the supernova explosion, becoming an optically faint stripped star \citep[e.g. ][]{Paczynski67,Gotberg+17,Yoon+17}. As a result, the system would appear as a single-line spectroscopic binary (SB1), with either a compact object \citep[e.g.][]{Langer+20} or a stripped star \citep[e.g.][]{Paczynski67,Pols+91,Irrgang+20} as the companion, depending on whether a supernova has occurred. 
In this case, making a rough estimation of what is the expected relative incidence of SB1 systems among the \SSEMoutliers{} when compared to the population of pre-interaction binary system with a dimmer, low-mass companion, is not so straightforward. 

Our observations show that 38\,$\pm$\,6\% of the \SSEMoutliers{} are runaways, compared to 26\,$\pm$\,3\% in the rest of the sample, supporting our hypothesis. Notably, we also find a higher incidence of SB1 systems among the \SSEMoutliers{}; 45\,$\pm$\,7\%, compared to 35\,$\pm$\,4\% in the rest of the sample. Further investigation into the companions of all detected SB1 systems in both subsamples could provide crucial insights to strengthen the likely post-binary interaction origin of the sample characterized by high He and unexpectedly low N surface abundances.

\subsection{A representative case: Cyg X-1}

Among the \SSEMoutliers{}, we find the well-known high-mass X-ray binary HD\,226868 (optical counterpart of Cyg X-1). It presents significant He enrichment (\He{}\,=\,0.18\,$\pm$\,0.05) but a clear deficiency in N (8.08\,$\pm$\,0.15\,dex) compared to the \SSEM{} predictions (\Nab{}\,$\sim$\,8.75\,dex, see Fig.~\ref{sum}).
This system hosts a black hole (BH) in a 5.6-day orbit, indicating that the visible component (an O9.7\,Iab star) likely received mass at some point in its evolution from the initially more massive component (i.e., the progenitor of the BH). This would be in line with the binary-evolution scenario we are proposing.

This information has clear implications for the evolutionary scenario previously assumed for the formation of this system \citep{Miller-Jones+21, Neijssel+21}, pointing to an important mass transfer phase of He-enriched material from the BH progenitor.

An abundance study of other known post-interaction candidates  \citep[e.g. Algols, Be stars, etc., see ][]{Sen+22, Dufton+24} could help us to confirm our hypothesis. Nonetheless, the abundance determination in these types of sources is still challenging.

\section{Final remarks}\label{Conc}

In this Letter, we presented reliable N and He surface abundances for a subset of 180 O-stars with \vsini{}\;$\leq\!150$~\kms{} and well-defined stellar parameters. 
Within this sample, we have found a subpopulation ($\sim\!20\%$ of the total sample and $\sim\!80\%$ of helium-enriched stars) of interesting objects with a surface enrichment pattern in He and N. This group is unexplainable within the framework of single-star evolutionary models and cannot be associated with any observational or analytical biases (see further notes in Appendix~\ref{appSingle}).

To explain this subset, we propose that its members are products of mass transfer in binary systems. Recent binary evolution simulations support this hypothesis by qualitatively reproducing our results (see Appendix~\ref{app_acc}), and can naturally explain the moderate N enrichment in strongly He-enhanced O stars. The higher prevalence of possible post-interaction outcomes (SB1 and runaways) among this subgroup supports this hypothesis. 
As a consequence of the proposed evolutionary scenario for these targets, the companions of the SB1 systems among them are promising candidates for stripped stars or black holes.

This study offers a critical observational reference for constraining key parameters of mass-transfer physics in close binary systems, particularly mass-transfer efficiency. Continued theoretical and observational efforts are crucial to deepen our understanding of massive star evolution, be it in single or multiple star systems.


\begin{acknowledgements}
We thank our anonymous referee for a set of useful comments and suggestions.
This research acknowledges the support from the State Research Agency (AEI) of the Spanish Ministry of Science and Innovation and Universities (MCIU) and the European Regional Development Fund (FEDER) under grant PID2021-122397NB-C21. 
This publication made use of the IAC HTCondor facility (http://research.cs.wisc.edu/htcondor/), partly financed by the Ministry of Economy and Competitiveness with FEDER funds, code IACA13-3E-2493.
C.M.S. acknowledges the workshop "Writing and Communicating your Science", organised by the Severo Ochoa Training Programme of the IAA-CSIC and imparted by Henri Boffin (ESO) and Johan Knapen (IAC) on 4-8 November 2024.
He also acknowledges Mar Carretero-Castrillo and Michelangelo Pantaleoni for kindly sharing their results on the runaway status of O Galactic stars for the statistical characterization of the sample.
H.J. received financial support for this research from the International Max Planck Research School (IMPRS) for Astronomy and Astrophysics at the Universities of Bonn and Cologne. The authors gratefully acknowledge the granted access to the Bonna cluster hosted by the University of Bonn.
Z.K. acknowledges support from JSPS Kakenhi Grant-in-Aid for Scientific Research (23K19071), the Overseas Visit Program for Young Researchers from the National Astronomical Observatory of Japan and the Early-Career Visitor Program from the Instituto de Astrofísica de Canarias. Numerical computations in part were carried out on the PC cluster at the Center for Computational Astrophysics, National Astronomical Observatory of Japan. 
J.P. acknowledges support from the Fundación Occident and the Instituto de Astrofísica de Canarias under the Visiting Researcher Programme 2022-2024 agreed between both institutions.

\end{acknowledgements}

\bibliographystyle{aa}
\bibliography{references}

\begin{appendix} 

\section{Stellar evolution models}\label{appMESA}

In this Appendix, we outline some of the main characteristics of the two grids of solar metallicity, single-star evolutionary models used to compare with our observational data. Additionally, Table~\ref{in_ab} displays the initial set of C, N, and O abundances considered by each of the models.

\subsection{Ekstr\"om et al. models}

\cite{Ekstrom+12} used the Geneva stellar evolution code, GENEC, to compute the model grids utilized in the current Letter. 
Convection is assumed as an instantaneous mixing process and $\alpha_{\rm MLT} = 1.6$ is adopted. A step overshooting with $\alpha_{\rm ov} = 0.1$ is used, leading to a radial extension of the nominal convective core size by 10 percent of the local pressure scale height. 
The initial abundances of hydrogen, helium, and metals are adopted as $X = 0.720$, $Y = 0.266$, and $Z = 0.014$, respectively. Initial abundances from C, N, and O can be found in Table~\ref{in_ab}. 
Rotational mixing is considered via diffusion coefficients that account for effective diffusivity (combining meridional circulation and horizontal diffusion) and shear diffusion.  
Angular momentum transport follows an advecto-diffusive scheme, in which meridional currents can increase the radial differential rotation and thus shears can become the dominant term for chemical mixing.
Rotation is assumed with $v/v_{\rm crit} = 0, 0.2, 0.4$. The critical velocity is defined as the velocity at which the gravitational acceleration is exactly counterbalanced by the centrifugal force; $v_{\rm crit}=\sqrt{\frac{2GM}{3R_{\rm pb}},}$ where $R_{\rm pb}$ is the polar radius at the critical limit.
The mass of the grid extends from 0.8 to 120~M$_\odot$. 

\subsection{Keszthelyi et al. models}

\cite{Keszthelyi+22} used the community-driven, open-source Modules for Experiments in Stellar Astrophyscs, MESA, software \citep[e.g.,][]{Paxton+15}. 
Convective mixing is considered a diffusive process with an efficiency of $\alpha_{\rm MLT} = 1.8$, and exponential overshooting is adopted with $f_{\rm ov} = 0.025$ and $f_{0} = 0.005$, which would roughly correspond to $\alpha_{\rm ov} = 0.2$. 
For their solar metallicity models, the initial abundances of $X,Y,Z$ are the same as in the \cite{Ekstrom+12} grid. The initial C, N, and O abundances are obtained from the determination of \cite{Nieva&Pryzbilla12} and thus the values are slightly different used by \cite{Ekstrom+12} (see first and third column in Table~\ref{in_ab}).

Due to the uncertain nature of chemical mixing, two schemes are implemented. "Mix1" follows the typical MESA approach, utilizing the prescriptions developed by \cite{Pin+89}, and using scaling factors to mitigate the otherwise too efficient mixing (compared to angular momentum transport). The "Mix2" scheme is the implementation of the \cite{Zahn92} equations, similar to the study of \cite{Ekstrom+12}.
Angular momentum transport follows a diffusive approximation. In this case, the Eddington-Sweet term is the dominant one for both angular momentum transport and chemical mixing. 
For this reason, the Mix2 models are still considerably different than those of \cite{Ekstrom+12}, since the underlying rotation profiles are distinct. In fact, the Mix2 models lead to quasi-chemically homogeneous evolution, which is the most efficient form of mixing in a star.
Rotation is assumed with an initial value of $\Omega/\Omega_{\rm crit} = 0.5$, corresponding to $\approx 350$~km\,s$^{-1}$. 
The initial mass range of the grid is from 3 to 60~M$_\odot$. 
For this study, it was beneficial to extend the parameter space of this grid, for which we computed new models.
These include initial masses of 80, 90, and 100~M$_\odot$, and initial rotation values of $\Omega/\Omega_{\rm crit} = 0.1, 0.2, 0.3, 0.4$ were additionally considered for the initial mass range of  20 - 100 M$_\odot$.

\section{Potential single-star evolution origin of the \SSEMoutliers{}}\label{appSingle}

In this Appendix, we summarize the main outcomes from our study exploring a potential single-star evolutionary origin of the \SSEMoutliers{}.

\subsection{Potential errors and biases in the spectroscopic analysis}\label{obs_errors}

Initially, we examined whether potential analysis errors could affect the accuracy of our He and N abundance measurements. In particular, we took a conservative approach when addressing formal uncertainties, considering the impact of EW measurements, the scatter from individual lines, stellar parameters, and microturbulence. We also performed a MCMC analysis to identify the \SSEMoutliers{} based on their distribution in the \He{}-\Nab{} abundance diagram. Furthermore, we carefully defined the boundaries of the blue open contour around the \SSEMoutliers{} to ensure that, even with uncertainties, these stars could not be explained by single-star evolution models.

It is also unlikely that a bias in the N abundance determination could explain the predominantly low values observed among the \SSEMoutliers{}. Any inherent observational bias would likely skew measurements toward higher \Nab{} values, as lower abundances result in fainter spectral lines that are harder to measure, potentially leading to the exclusion of such stars. In contrast, He exhibits stronger lines, making a high-abundance bias negligible. However, dilution effects from external contamination in the continuum could slightly underestimate He abundance.

\subsection{Gravity darkening effects}\label{3D_effects}

In \SSEM{}, rotational mixing is expected to be the primary driver of surface contamination by CNO-cycle products. For O-type stars, He enrichment (i.e., \He{}\,$\gtrsim$\,0.12) is expected to be detectable only in stars with initial masses above  $\sim~30~{\rm M}_{\odot}$ and equatorial rotational velocities at birth exceeding 300~\kms{} \citep[e.g.,][see also Fig.~\ref{sHR}]{Ekstrom+12}. While significant braking of the stellar surface is predicted to occur during the first half of the main sequence evolution under certain conditions \citep[see, e.g.,][]{Ekstrom+12, Keszthelyi+22, Holgado+22}, not all \SSEM{} scenarios result in this outcome \citep[e.g.,][]{Brott+11}. Consequently, some stars in our sample may still be fast rotators with relatively low inclination angles (as we are limited to \vsini\ $\leq$ 150 \kms).

In this context, as noted by \citet{Fremat2005} and \citet{Abdul-Masih23}, neglecting the 3D deformations caused by rapid rotation, including the effects of gravity darkening \citep{von_Zeipel24}, can significantly impact stellar parameters and abundance determinations. \cite{Abdul-Masih23} highlights a temperature-dependent discrepancy in helium abundance that could notably deviate from the actual value.
Furthermore, potentially similar effects in nitrogen warrant an investigation. However, the probability that all \SSEMoutliers{} are fast rotators observed at near pole-on orientations is negligible (e.g., \citealt{Holgado+22}).

\subsection{Impact of baseline abundances}\label{subsolar} 

We investigated the influence of the baseline abundances in the interpretation of our results. In particular, we studied the possibility that some of the \SSEMoutliers{} correspond to stars with an initial CNO composition lower than solar.

\begin{figure}[!t]
\includegraphics[width=1.\hsize,trim={0 15 0 0}]{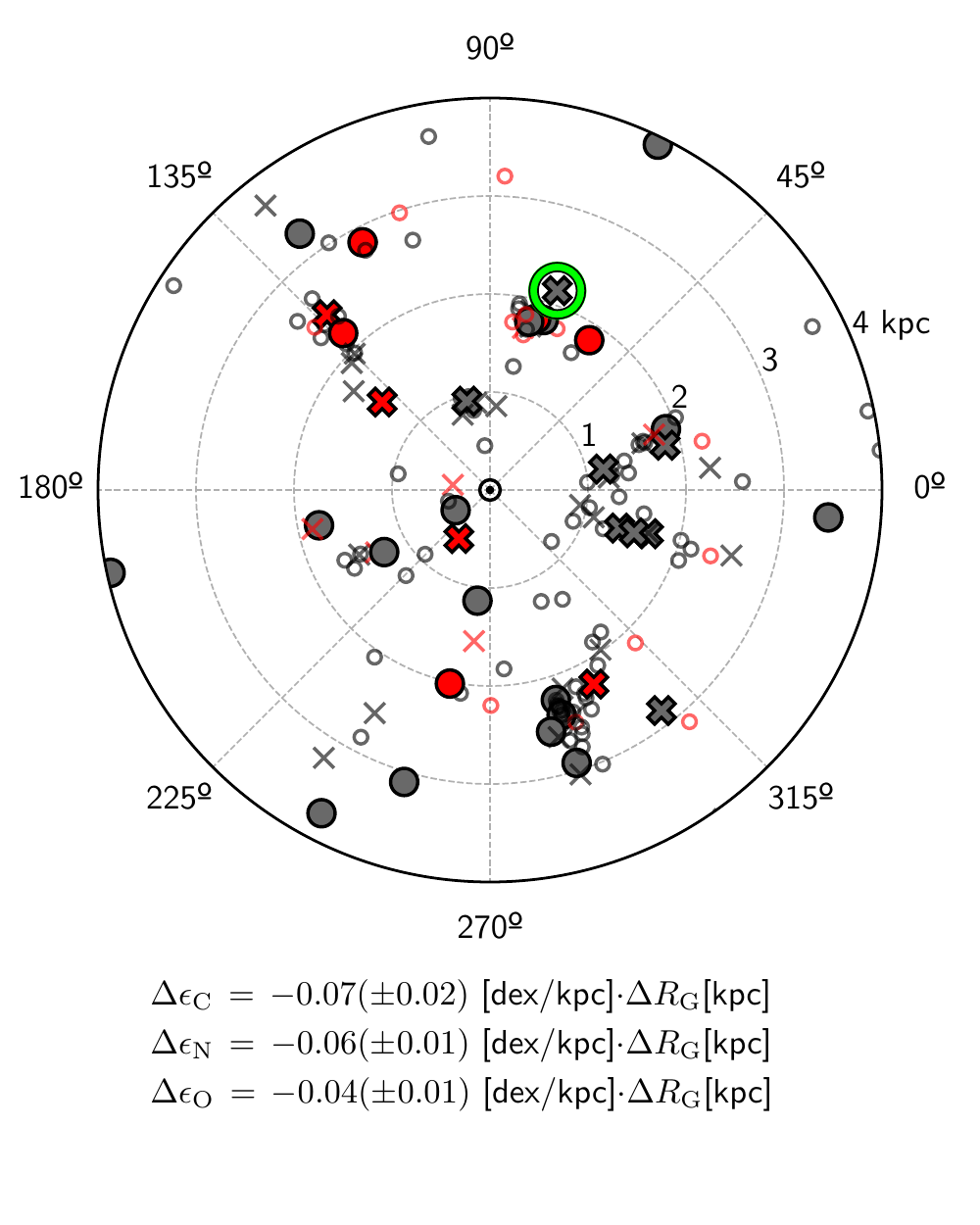}
  \caption{
  {\it Top:} Spatial distribution of our sample of O-type stars in the Galactic plane,  with the Sun ($\odot$) at the center. 
  Same marker code as in Fig.~\ref{sum}. 
  {\it Bottom:} Values of the Galactic present-day abundance gradient of C, N, O as a function of galactocentric distance (R$_{\rm G}$), from \cite{Arellano-Cordova+20, Arellano-Cordova+21}. 
 Distances from \cite{Bailer-Jones+21}
 \label{gal_map}}
\end{figure}

Our stars reach distances up to $\sim$\,4\,kpc from the Sun (Fig.~\ref{gal_map}).
However, this distribution is not homogeneous. Most stars are located within approximately 3\,kpc, with this distance decreasing to around 2\,kpc in the direction of the Galactic anticenter.
Overall, the initial abundances in our sample are affected by the Galactic chemical abundance gradient. \cite{Arellano-Cordova+20, Arellano-Cordova+21} investigated the present-day gradient of the Galactic disc as delineated by H\,{\sc ii} regions.
Together with the Cosmic Abundance Standard (CAS) from \cite{Nieva&Pryzbilla12}, it gives us a rough estimate of the minimum expected initial C, N, and O abundances for our sample of $\sim$8.1, 7.6, and 8.6\,dex, respectively.
This could explain the existence of stars with N abundances below CAS in Fig.~\ref{sum} (shaded brown rectangle).

For the comparative purpose of this section, we found the \SSEM{} of \cite{Eggenberger+21} to be the most appropriate.
Those models were computed with the GENEC code for a metallicity $Z$\,$\sim$\,0.45\,$Z_{\odot}$ 
(i.e., a value typically considered for the LMC).
For the initial abundances of elements heavier than He, they scaled the value from \cite{Ekstrom+12} with metallicity (Table~\ref{in_ab}).
The resulting initial CNO composition is below the minimum expected abundances inferred from the gradient of the Milky Way disc. Moreover, it reaches the lowest end of our N abundance distribution in Fig.~\ref{sum}.

\begin{table*}
\label{in_ab}
\centering
\caption{Initial abundances of the models considered.}
\begin{tabular}{ccccc}
\hline \hline
& \cite{Ekstrom+12} & \cite{Eggenberger+21}  & \cite{Keszthelyi+22} & \cite{Jinprep} \\
\hline
$\epsilon_{\rm C, i}$ [$X_{\rm C, i}/10^{-3}$] & 8.39 [2.31] & 8.01 [0.99] & 8.33 [1.85] & 8.52 [2.85]\\
$\epsilon_{\rm N, i}$ [$X_{\rm N, i}/10^{-3}$] & 7.78 [0.66] & 7.40 [0.28] & 7.79 [0.62]  & 7.88 [0.77] \\
$\epsilon_{\rm O, i}$ [$X_{\rm O, i}/10^{-3}$] & 8.66 [5.73] & 8.28 [2.46] & 8.76 [6.63]  & 8.74 [6.39] \\
$X_{\rm CNO, i}/10^{-3}$ & 8.7 & 3.73 & 9.1 & 10.01 \\
\hline
\end{tabular}
\tablefoot{Initial C, N, and O abundances assumed in the \SSEM{} considered in this work.  Individual C, N, O abundances in the usual $\epsilon_{\rm X}\,=\,\log(N({{\rm X}})/N({{\rm H}}))\,+\,12$ scale, and in brackets in mass fraction. Total CNO ($X_{\rm CNO, i}$) abundances in mass fraction.
In \citeauthor{Eggenberger+21}, initial abundances of elements different than H and He were scaled with metallicity (see text).}
\end{table*}

As the main result of this exercise, we found that when we consider such an extreme case of lower baseline abundances -- inappropriate for our sample --, we could account for some of our problematic sources. Still, even in this extreme scenario, about 10\% of the analyzed stars would retain their status as \SSEMoutliers{} (see stars surrounded by the orange contour in Fig~\ref{sum}). 
Moreover, a similar argument based on the Galactic gradient could lead to more sources closer to the Galactic Center appearing as \SSEMoutliers{}.

\subsection{\SSEMoutliers{} in the spectroscopic HR diagram}

\begin{figure}[!t]
\includegraphics[width=1.1\hsize]{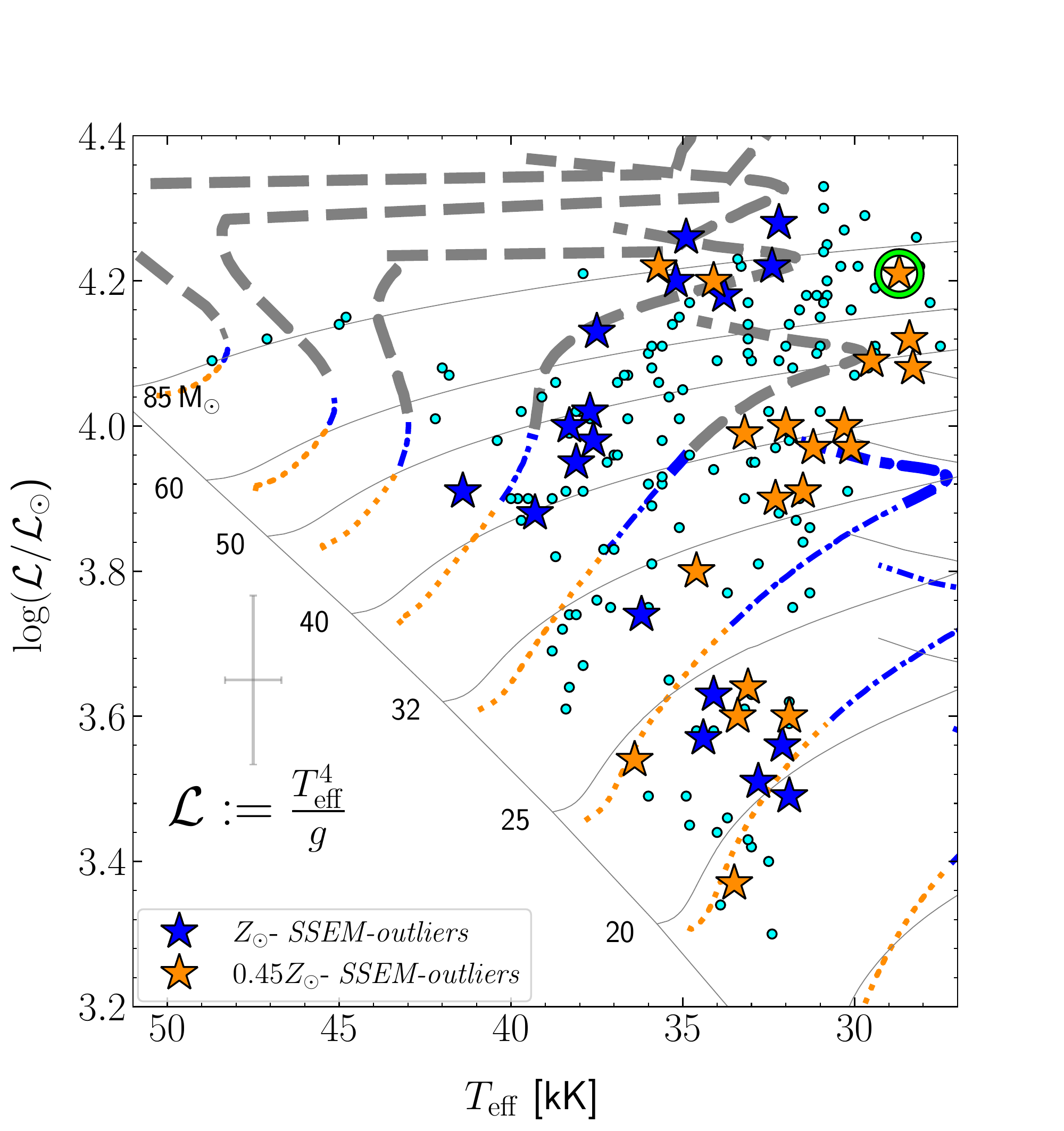}
  \caption{
  Spectroscopic HRD of the 180 Galactic O-type stars studied, with \SSEMoutliers{} marked by star symbols. Orange stars indicate those that remain outliers even when assuming \cite{Eggenberger+21} models (subsolar baseline abundances). HD\,226868, the optical counterpart of Cyg X-1, is circled in green. Non-rotating and $v_{\rm ini}/v_{\rm crit}=0.4$ evolutionary tracks from \cite{Ekstrom+12} are shown for reference (see text for details). Typical uncertainties in \Teff\ and \Lspalone\ are represented by a gray cross.}
     \label{sHR}
\end{figure}

Figure~\ref{sHR} shows the location of the 180 Galactic O-type stars from our sample in the spectroscopic HR diagram \citep[sHRD, introduced by][]{Langer14}. \SSEMoutliers{} are marked with star symbols. Among these, those explainable by \SSEM{} with subsolar baseline abundances are shown in dark blue, while the remaining outliers, even in this extreme case, are highlighted in orange. The latter group clusters in the cooler region of the diagram ($30 \lesssim T_{\rm eff}/{\rm kK} \lesssim 35$), whereas the former spans a wider temperature range ($32 \lesssim T_{\rm eff}/{\rm kK} \lesssim 43$).

The figure also includes the \SSEM{} tracks from \cite{Ekstrom+12}, both without rotation (solid lines) and with an initial rotational velocity of 40\% of the critical value (dashed lines). For the rotating models\footnote{Non-rotating models do not produce remarkable surface enrichment during core hydrogen burning.}, we divide the tracks into three regions based on the surface N abundance, following the behavior of evolutionary models in Fig.~\ref{sum}. The regions are marked by an orange-dotted line, a blue dashed-dotted line, and a gray dashed line, representing \Nab{}\,$\leq$\,8.25~dex, 8.25\,$<$\,\Nab{}\,$\leq$\,8.6~dex, and \Nab{}\,$>$\,8.6~dex, respectively. The 8.25 value represents the maximum N enrichment in subsolar abundances models, while 8.6 corresponds to the solar-abundance models. 
In addition, we highlight (by thick lines) the regions of the tracks where \He{}$\geq0.12$. 
Remarkably, this value is not reached during the MS for initial masses below $\sim$30\,$M_{\odot}$, and is only reached after the stars exhibit CNO-equilibrium (\Nab{}$>$8.6\,dex).

The overall agreement between observational data for the \SSEMoutliers{} and predictions from the rotating models is very weak. A significant portion of the sample (to repeat in different units: particularly stars with log($\mathcal{L}$/$\mathcal{L}_{\odot}$)\,$\lesssim$\,3.8\,dex, corresponding to the afore-mentioned tracks below $\sim$30\,$M_{\odot}$) is not expected to reach \He{}\,$>$\,0.12 during the MS. While higher-mass stars are predicted to show some helium enrichment, this is never consistent with the low N abundances observed in the objects marked with star symbols (see Table~\ref{BinList_table}).

Finally, efficient rotational mixing should lead to increased N at the surface when stars evolve away from the ZAMS. Consequently, those objects marked with orange symbols (i.e., \SSEMoutliers{} with lower N abundances) would be expected to lie closer to the ZAMS than the blue-marked \SSEMoutliers{}. Their distribution should roughly overlap with the region of the rotating tracks highlighted in orange. However, Fig.~\ref{sHR} shows the opposite trend, complicating the interpretation of these \SSEMoutliers{} as products of single-star evolution.

\section{Comparing models of mass gainers and of single stars}\label{app_acc}

\def\Y{\mathrm{(He/H)}}  

\begin{figure*}
    \centering
    \includegraphics[width=0.49\textwidth]{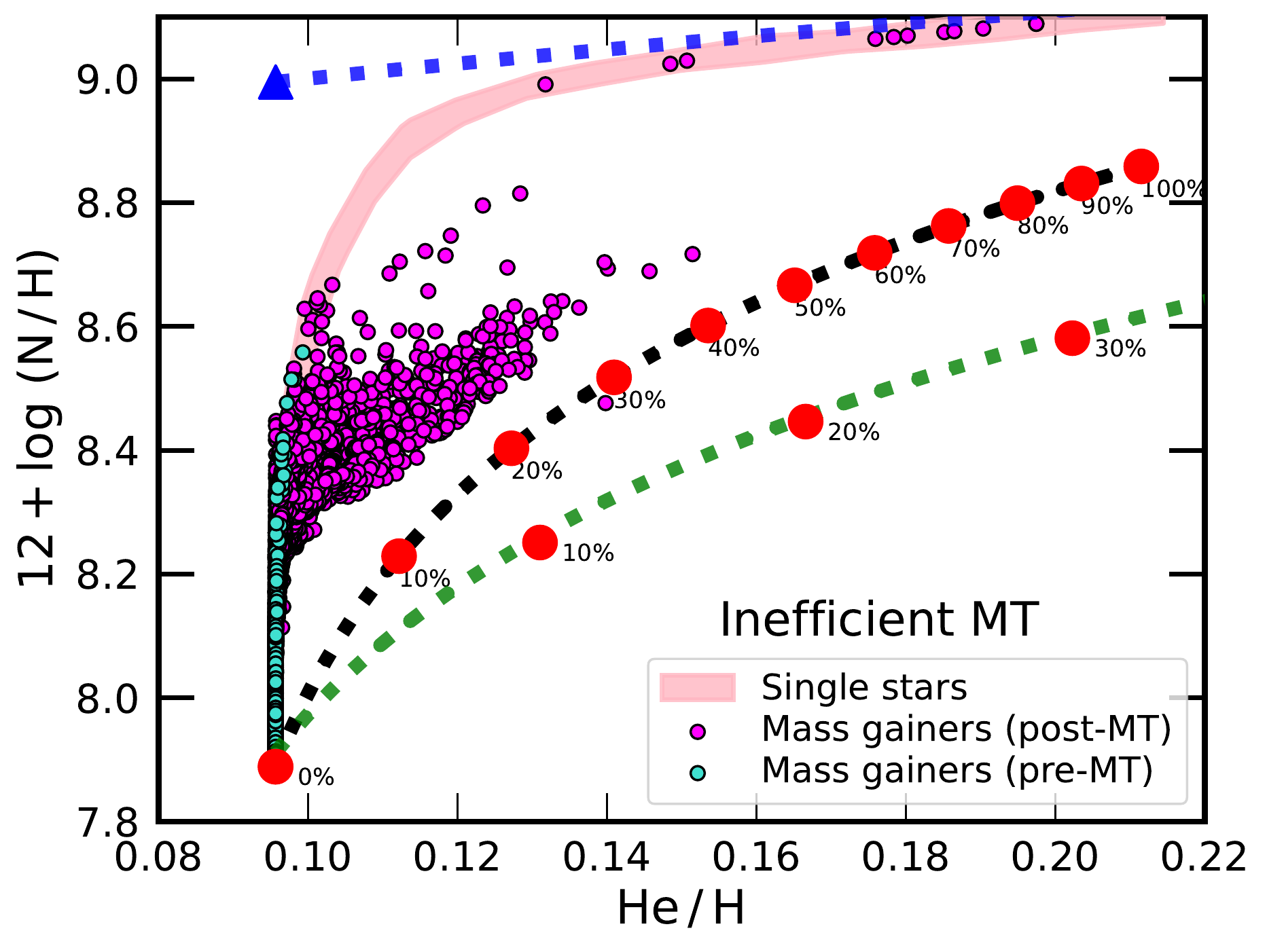}
    \includegraphics[width=0.49\textwidth]{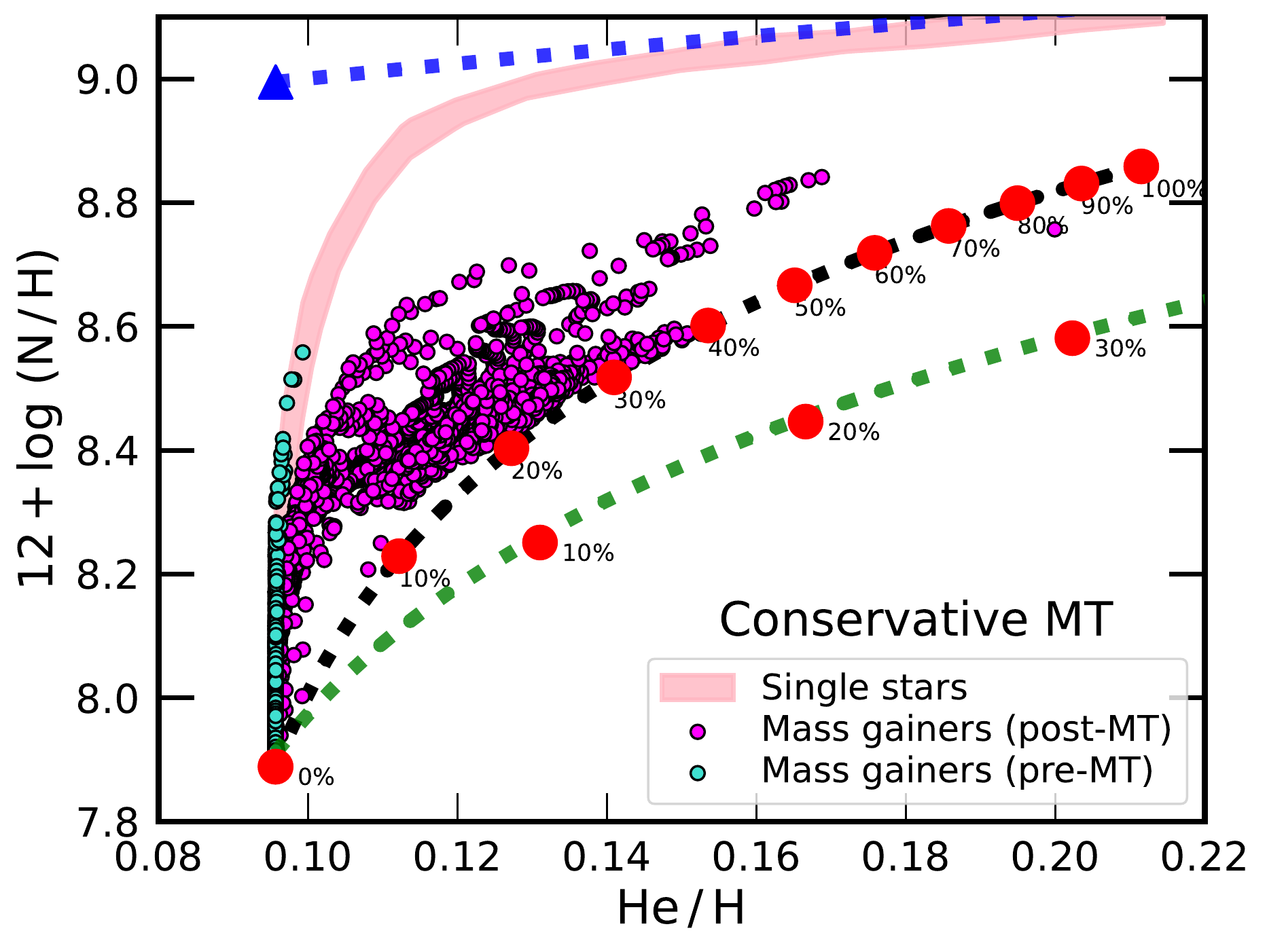}
\caption{Location of single and binary evolutionary models from \citet{Jin+24,Jinprep} in the N\,vs.\,He diagram. The extended pink line covers the evolutionary tracks of the \SSEM{} described in Appendix~\ref{app_acc}. Mass gainers predictions are depicted immediately before mass transfer (MT; cyan dots), and after mass transfer at a time when thermal equilibrium is restored (magenta dots). Left and right panels depict MESA calculations assuming an inefficient or conservative mass transfer, respectively. The dotted-blue line shows abundance ratios under the assumption of complete CNO-equilibrium at any given He abundance. The dotted-green line has the same meaning but assuming a mixture between matter with the initial chemical composition and matter consisting of pure He and the CNO-equilibrium value of N (with a mass fraction indicated by along the red dots). Both dotted lines frame the allowed space for evolutionary models. The dotted-black line refers to a mixture of matter with the initial abundances with matter with a He mass fraction of $Y=0.63$, corresponding to the average in the He-enriched part of the donor envelopes, and the CNO-equilibrium value of N.
\label{acc}
} 
\end{figure*}

Figure~\ref{acc} compares the surface He and N abundance predictions from detailed models of single stars and mass gainers in 
binary systems. Both sets of models were computed with MESA under similar physical assumptions \citep{Jin+24,Jinprep}. Initial abundances in all models are based on the protosolar values from \citet{Asplund2021}, which differ from those used in the \SSEM{}  by \citet{Ekstrom+12} and \citet{Keszthelyi+22} (Table~\ref{in_ab}).

The single star models cover an initial mass range of $M_\mathrm{i}$\,=\,10\,--\,100\,$M_\odot$. Models with $M_\mathrm{i}$\,=\,10\,--\,40\,$M_\odot$ are taken from \citet{Jin+24}, while those for higher masses have been coherently computed afterwards. In all these models, initial rotation was set to $v/v_{\rm crit} = 0.4$

Mass gainer predictions are taken from a comprehensive grid of detailed binary evolution models consisting of $\sim$38\,000 initial configurations. These cover primary star masses in the range 5\,--\,100\,$M_\odot$, mass ratios of 0.1\,--\,0.95, and orbital periods ranging from that resulting in Roche-lobe filling at zero-age main sequence (ZAMS) to non-interaction. In all cases, the initial rotation rate was set to $v_\mathrm{i}$/$v_\mathrm{crit}$\,=\,0.2.

In order to match the evolutionary masses of the observed stars (Fig.~\ref{sHR}), the two panels in Fig.~\ref{acc} only depict predictions for mass gainers with $M$\,$>$15\,$M_\odot$.
Each mass gainer is represented by a single point in the N\,vs.\,He diagrams. In particular, the considered abundances correspond to the moment when the mass gainer has thermally relaxed after the mass transfer event and thermohaline mixing of the enriched matter has taken place.
Thermohaline mixing is treated as a diffusive process in MESA, with a parameter $\alpha_{\rm th}$ determining the speed of mixing. We adopt $\alpha_{\rm th}=1$ \citep{Kippenhahn+80}. Thermohaline mixing occurs on the thermal timescale \citep{Cantiello&Langer10}. Consequently, the envelope of the mass gainer will achieve full mixing on a short timescale compared to the post-interaction MS lifetime.
After this phase, the surface abundances change minimally.  

In the original calculations by 
\textcolor{blue}{\cite{Jinprep}}, 
the amount of accreted mass was limited by the rotation of the mass gainer. When spin-down due to tides is inefficient, mass transfer is highly non-conservative, and mass gainers end up accreting only a few percent of the transferred mass \citep{Langer12}. However, the mass transfer physics and its efficiency are not well-constrained, and there are several pieces of evidence supporting a higher mass transfer efficiency \citep[e.g.,][]{Wellstein&Langer99, deMink+07, Schootemeijer+18, Vinciguerra2020}. Thus, in addition to exploring the original mass gainer models (left panel of Fig.~\ref{acc}), we recompute the chemical envelope structure of the mass gainers assuming fully conservative mass transfer (right panel).

For the conservative mass transfer case, we estimated the composition of the mass gainer by assuming that all the transferred mass is accreted onto the mass gainer and, subsequently, the polluted envelope is fully mixed with the original material.
The accreted matter is mixed down partly into the H/He composition gradient such that the mean molecular weight of the fully mixed matter in the envelope is the same as that of the layer in the H/He composition gradient. This is expected when thermohaline mixing has fully taken place.
Noteworthy, we consider only binary models which are expected to survive the conservative mass transfer. To this aim, we compare the thermal timescale and the mass gain timescale of the mass gainer following \citet{Schuermann2024}. In particular, we assume $\dot{M_\mathrm{2}} = \dot{M}_\mathrm{1,tr} = M_\mathrm{env,1} / \tau_\mathrm{KH,1}$ when considering their eq.\,5, where $\dot{M_\mathrm{2}}$ is the change of the mass per unit time of the mass gainer, $\dot{M}_\mathrm{1,tr}$ is the mass transfer rate from the donor, $M_\mathrm{env,1}$ and $\tau_\mathrm{KH,1}$ are the envelope mass and the thermal timescale of the donor. As for the inefficient mass transfer case, we only depict in Fig.~\ref{acc} (right figure) those mass gainers with $M > 15 M_\odot$. 

Mass gainers will become fast rotators and rotational mixing will be enhanced. However, the mean molecular weight gradient already established above the core will likely prevent the inner nuclear-processed matter (He-rich, N-rich) from being transported to the surface \citep[see discussion in Sect. 6.2 of][]{Dufton+24}.
Moreover, as we consider the full mixing of the accreted matter, we do not expect rotational mixing to further affect surface abundances. 
Only under the scenario of the mass gainer already fast rotating before the mass accretion, the mass gainers would show surface N-enhancement closer to the CNO equilibrium value. However, this is unlikely when assuming that the initial rotation rates of main sequence binary components are following the same distribution as those of single stars, as \cite{Ramirez-Agudelo+15} found for O star binaries in the LMC.

As shown in Fig.~\ref{acc}, for a given He enrichment, the mass gainer models display a much weaker N enrichment than the single star models. In the latter, mixing processes can bring N much easier to the surface than He, a fact which is irrelevant for the mass gainers. While conservative mass transfer can lead to higher He enrichments than inefficient mass transfer, the offset of the mass gainer models from the single-star models is evident in both cases.

For reference purposes, we indicate in both panels of Fig.~\ref{acc} 
the extreme cases in between which any stellar evolution model must lie\footnote{For simplicity, in both considered cases we assume that all the CNO isotopes convert into $^{14}\mathrm{N}$.}. The blue-dotted line assumes CNO-equilibrium is reached for any given He abundance
$${\rm N}/{\rm H} = X_\mathrm{CNO,i}\,(1+4\,\Y)\,/\,14,$$ 
while the green-dotted line is obtained by mixing pure He matter, which is in CNO equilibrium, 
$${\rm N}/{\rm H} = \frac{X_\mathrm{N,i}+X_\mathrm{CNO,i\,}(4\,(1-Y_\mathrm{i})\,\Y-Y_\mathrm{i})}{14\,(1-Y_\mathrm{i})},$$
where He/H and N/H represent number fractions, while
$Y_\mathrm{i}$, $X_\mathrm{N,i}$, and $X_\mathrm{CNO,i}$ indicate initial mass fractions of He, N, and C+N+O, respectively. 
These reference limits have been taken into account for the discussion presented in Sect.~\ref{evolpred} (see also Fig.~\ref{sum}).

\section{Stellar parameters for \SSEMoutliers{}}

Table~\ref{BinList_table} summarizes observational information of interest about the 36 \SSEMoutliers{} surrounded by a blue contour in Fig.~\ref{sum} and highlighted with larger, bold symbols.  

\begin{table*}
\caption{Stellar parameters of the \SSEMoutliers{}.}
\label{BinList_table}
\centering
\begin{tabular}{ccccccccc}
\hline \hline
Star & SpC & \Teff{} & ${\rm log}(\mathcal{L}/\mathcal{L}_{\odot})$ & \vsini{} & \He{} & \Nab{} & $\xi_{\rm N}$ & Status\\
\hline

BD\,+57\,247 & O9.7\,IV & 32.8\,$\pm$\,1.6 & 3.5\,$\pm$\,0.3 & 28 & 0.17\,$\pm$\,0.09 & 8.45\,$\pm$\,0.12 & 6 & MD\\
HD\,326329 & O9.7\,V & 31.9\,$\pm$\,0.4 & 3.49\,$\pm$\,0.05 & 87 & 0.14\,$\pm$\,0.02 & 8.15\,$\pm$\,0.12 & 4 & SB1\\
HD\,36512 & O9.7\,V & 33.5\,$\pm$\,0.7 & 3.37\,$\pm$\,0.14 & 13 & 0.12\,$\pm$\,0.03 & 7.84\,$\pm$\,0.12 & 4 & LS\\
HD\,38666 & O9.5\,V & 33.1\,$\pm$\,0.4 & 3.64\,$\pm$\,0.05 & 111 & 0.14\,$\pm$\,0.02 & 7.81\,$\pm$\,0.15 & 10 & SB1, RW\\
HD\,202214 & O9.5\,IV & 32.1\,$\pm$\,1.0 & 3.56\,$\pm$\,0.16 & 26 & 0.13\,$\pm$\,0.04 & 8.07\,$\pm$\,0.11 & 6 & LS\\
HD\,93027 & O9.5\,IV & 33.4\,$\pm$\,0.7 & 3.60\,$\pm$\,0.13 & 46 & 0.12\,$\pm$\,0.03 & 7.67\,$\pm$\,0.12 & 8 & MD\\
HD\,12323 & ON9.2\,V & 34.4\,$\pm$\,1.0 & 3.57\,$\pm$\,0.19 & 121 & 0.18\,$\pm$\,0.07 & 8.32\,$\pm$\,0.15 & 8 & SB1, RW\\
HD\,76341 & O9.2\,IV & 32.3\,$\pm$\,0.7 & 3.90\,$\pm$\,0.09 & 51 & 0.15\,$\pm$\,0.04 & 7.8\,$\pm$\,0.12 & 17 & LS\\
HD\,44597 & O9.2\,V & 34.1\,$\pm$\,1.1 & 3.6\,$\pm$\,0.2 & 15 & 0.12\,$\pm$\,0.04 & 7.95\,$\pm$\,0.12 & 9 & LS\\
HD\,14633 & ON8.5\,V & 34.6\,$\pm$\,0.8 & 3.8\,$\pm$\,0.13 & 121 & 0.19\,$\pm$\,0.07 & 8.18\,$\pm$\,0.20 & 15 & SB1, RW\\
HD\,48279 & O8.5\,V\,zNstrvar? & 36.2\,$\pm$\,0.7 & 3.74\,$\pm$\,0.12 & 131 & 0.14\,$\pm$\,0.02 & 8.24\,$\pm$\,0.15 & 13 & LS\\
ALS\,15196 & O8.5\,V & 36.4\,$\pm$\,0.9 & 3.54\,$\pm$\,0.13 & 66 & 0.12\,$\pm$\,0.03 & 7.81\,$\pm$\,0.12 & 6 & MD\\
HD\,90273 & ON7\,V\,((f)) & 38.3\,$\pm$\,0.7 & 4.00\,$\pm$\,0.08 & 55 & <0.16 & 8.39\,$\pm$\,0.12 & 20 & MD\\
HD\,193595 & O7\,V\,((f)) & 37.6\,$\pm$\,0.6 & 4.0\,$\pm$\,0.1 & 46 & 0.15\,$\pm$\,0.02 & 8.35\,$\pm$\,0.12 & 19 & LS\\
HD\,12993 & O6.5\,V\,((f)) Nstr & 39.3\,$\pm$\,1.0 & 3.88\,$\pm$\,0.15 & 84 & 0.16\,$\pm$\,0.05 & 8.40\,$\pm$\,0.15 & 21 & LS, RW\\
HD\,167633 & O6.5\,V\,((f)) & 37.7\,$\pm$\,0.9 & 4.0\,$\pm$\,0.1 & 129 & 0.13\,$\pm$\,0.04 & 8.28\,$\pm$\,0.15 & 21 & SB1\\
HD\,63005 & O6.5\,IV & 38.1\,$\pm$\,1.4 & 3.95\,$\pm$\,0.19 & 57 & 0.12\,$\pm$\,0.04 & 7.97\,$\pm$\,0.12 & 18 & MD\\
HD\,256725 & O5\,V((fc))z & 41.4\,$\pm$\,1.3 & 3.91\,$\pm$\,0.15 & 67 & 0.12\,$\pm$\,0.03 & 7.95\,$\pm$\,0.15 & 17 & LS\\

\hline
CPD\,-35\,2105 & O9.2\,III & 31.5\,$\pm$\,0.6 & 3.91\,$\pm$\,0.11 & 80 & 0.11\,$\pm$\,0.03 & 7.56\,$\pm$\,0.12 & 17 & MD\\
HD\,152247 & O9.2\,III & 31.2\,$\pm$\,0.5 & 3.97\,$\pm$\,0.11 & 82 & 0.18\,$\pm$\,0.05 & 7.67\,$\pm$\,0.15 & 19 & SB1\\
HD\,305523 & O9\,II-III & 32.0\,$\pm$\,0.4 & 4.0\,$\pm$\,0.05 & 57 & 0.12\,$\pm$\,0.03 & 7.67\,$\pm$\,0.12 & 20 & MD\\
HD\,105627 & O9\,III & 33.2\,$\pm$\,0.6 & 3.99\,$\pm$\,0.09 & 141 & 0.14\,$\pm$\,0.05 & 7.89\,$\pm$\,0.15 & 21 & SB1, RW\\
HD\,190864 & O6.5\,III\,(f) & 37.5\,$\pm$\,0.7 & 4.13\,$\pm$\,0.08 & 66 & 0.14\,$\pm$\,0.02 & 8.31\,$\pm$\,0.20 & 23 & LS\\
\hline

HD\,226868 & O9.7\,Iab\,pvar & 28.7\,$\pm$\,0.6 & 4.21\,$\pm$\,0.08 & 95 & 0.18\,$\pm$\,0.05 & 8.08\,$\pm$\,0.12 & 26 & SB1\\
HD\,225146 & O9.7\,Iab & 28.3\,$\pm$\,0.9 & 4.08\,$\pm$\,0.14 & 67 & 0.15\,$\pm$\,0.06 & 7.87\,$\pm$\,0.12 & 22 & LS, RW\\
HD\,152405 & O9.7\,II & 30.3\,$\pm$\,0.7 & 4.00\,$\pm$\,0.09 & 59 & 0.14\,$\pm$\,0.04 & 8.07\,$\pm$\,0.12 & 20 & SB1\\
HD\,75222 & O9.7\,Iab & 29.5\,$\pm$\,0.9 & 4.09\,$\pm$\,0.11 & 86 & 0.14\,$\pm$\,0.04 & 7.95\,$\pm$\,0.11 & 23 & LS, RW\\
HD\,167264 & O9.7\,Iab & 28.4\,$\pm$\,0.6 & 4.12\,$\pm$\,0.11 & 71 & 0.12\,$\pm$\,0.02 & 7.94\,$\pm$\,0.11 & 23 & SB1\\
HD\,36486 & O9.5\,II\,N\,wk & 30.1\,$\pm$\,0.5 & 3.97\,$\pm$\,0.06 & 100 & 0.11\,$\pm$\,0.02 & 7.70\,$\pm$\,0.12 & 19 & SB1\\
BD\,-11\,4586 & O8\,Ib\,(f) & 32.4\,$\pm$\,1.1 & 4.22\,$\pm$\,0.15 & 74 & 0.13\,$\pm$\,0.05 & 8.12\,$\pm$\,0.12 & 26 & MD\\
HD\,188001 & O7.5\,Iab\,f & 32.2\,$\pm$\,0.4 & 4.28\,$\pm$\,0.05 & 69 & 0.23\,$\pm$\,0.02 & 8.58\,$\pm$\,0.28 & 28 & LS, RW\\
HD\,332755 & O7.5\,Ib-II & 35.2\,$\pm$\,1.6 & 4.20\,$\pm$\,0.22 & 2 & 0.15\,$\pm$\,0.06 & 8.21\,$\pm$\,0.13 & 26 & LS\\
HD\,192639 & O7.5\,Iab\,f & 34.1\,$\pm$\,0.5 & 4.20\,$\pm$\,0.05 & 82 & 0.14\,$\pm$\,0.02 & 8.04\,$\pm$\,0.16 & 26 & LS, RW\\
HD\,117797 & O7.5\,fp & 33.8\,$\pm$\,0.5 & 4.18\,$\pm$\,0.05 & 150 & 0.14\,$\pm$\,0.02 & 8.22\,$\pm$\,0.15 & 26 & SB1\\
HD\,69464 & O7\,Ib\,(f) & 35.7\,$\pm$\,1.0 & 4.22\,$\pm$\,0.11 & 73 & 0.14\,$\pm$\,0.03 & 8.05\,$\pm$\,0.28 & 26 & LS\\
HD\,163758 & O6.5\,Ia\,fp & 34.9\,$\pm$\,0.3 & 4.26\,$\pm$\,0.06 & 76 & 0.20\,$\pm$\,0.02 & 8.50\,$\pm$\,0.13 & 27 & LS\\
\hline
\end{tabular}
\tablefoot{Stellar parameters for the 36 \SSEMoutliers{}. Effective temperature (in kK), spectroscopic luminosity, He (as number ratio) and N abundances along with their computed uncertainties. In the case of \vsini\ (in km\,s$^{-1}$), it has been considered a characteristic uncertainty of $10\%$ of the central value. Microturbulence ($\xi_{\rm N}$) in km\,s$^{-1}$. The last column indicates the binarity status as likely single (LS), single line spectroscopic binary (SB1), or undetermined due to insufficient spectra (MD: more data needed), and whether they are claimed to be runaways (RW) either in \cite{Maiz-Apellaniz+18} or \cite{Carretero-Castrillo+23}.
Spectral type (SpC) from the Galactic O-star catalog \citep[GOSC, ][]{Maiz-Apellaniz+13}.
Stars are sorted by luminosity class (LC) and then by increasing spectral type.
}
\end{table*}

\end{appendix}

\end{document}